\documentstyle[12pt]{article}

\author{R.H. Dalitz\\
Department of Physics (Theoretical Physics)\\
University of Oxford\\
1 Keble Road, Oxford OX1 3NP  UK
\and
Gary R. Goldstein\\
Department of Physics\\
Tufts University\\
Medford, MA 02155  USA}
\title{Top Mass Analyses for the Reported Top-Antitop Production and
Decay Events}
\date{ 2 June 1995}

\begin{document}
\maketitle
\begin{abstract}
We have analyzed the available data on $p\bar{p}\,\rightarrow\, t\,
\bar{t}$ followed by the decays ($t\rightarrow bW^{+},\:
\bar{t}\rightarrow \bar{b}W^{-}$) which lead to $e^{\pm}\mu^{\mp}2
\mathrm{jets}$ or $l^{\pm}4 \mathrm{jets}$ configurations, using a
likelihood method we proposed earlier. The outcome is compared with the
recent CDF analysis. In an appendix, we discuss the nature of the
additional ``slow'' $\mu^{+}$ observed in one CDF dilepton event,
concluding that it is most probably a ``tertiary lepton'' resulting from
the decay sequence $b\rightarrow c +\mathrm{hadrons}$, followed by
$c\rightarrow s \mu^{+} \nu$.
\end{abstract}
\newpage

\section{INTRODUCTION}

    A major step forward in research on the top quark has recently been
achieved by the CDF Collaboration\cite{cdf95} and the D0
Collaboration\cite{d095}. Both groups present evidence for the discovery
of the heaviest quark. Previously CDF\cite{cdf1,cdf2} published some  of
the data on a number of top-antitop production and decay events
\cite{cdf1,cdf2}. These
data consist of seven single lepton events with four large $p_T$ jets
($l^{\pm}4j$) and two opposite-sign dilepton events with two large
$p_T$ jets ($\mu^{\pm}e^{\mp}2j$)\footnote{These last two events each
have three jets, but the third jet has very high momentum with low $p_T$
and must be regarded as due to initial-state inner-bremsstrahlung. In
both events the third jet happens to be associated with the antiproton
(see Table I).}.
Each of the former has one jet ``tagged'' as a likely $b$ or
$\bar{b}$ jet and passes the many different cuts designed to minimize
the contribution from the extensive background \cite{berends}
 expected from ``W + jets'' Standard Model processes.
The dilepton events are fewer, since
hadronic decays predominate over leptonic decays for the W-boson;
indeed, for 7 b-tagged $l^{\pm}4j$ events
\footnote{The overall
efficiency of b-tagging \cite{cdf2} is about $0.22$ \cite{cdf3}.},
we would expect \cite{dalitz1}
about $7/(18\times 0.22) \approx 1.8 \mu^{\pm}e^{\mp}2j$ events,
compatible with observation. The CDF analysis of the
$l^{\pm}4j$ events follows established procedures from bubble chamber
physics, except that energy measurements by calorimeter have much less
percentage error than momentum measurements by magnetic fields. The CDF
analysis of the $\mu^{\pm}e^{\mp}2j$ events is only kinematical, based
on whether the measured momenta satisfy a set of cuts deemed sufficient
to remove all background. The formula for this rate as a function of top
mass $m_t$ is then used to deduce a lower limit on $m_t$. This
procedure gives no weight to precise detailed features of the event and their
relative probabilities, given our knowledge of the quark structure of
protons and of the Electroweak decay amplitudes characteristic of the
Standard Model, so well confirmed elsewhere today.

  It appears worthwhile at present to make independent calculations
from these and other data \cite{sliwa,cdf3,d0} by methods
we have already described, as a
contribution to the phenomenology of top quarks. This may also help
us to assess the relative benefits of the various methods which have been
proposed. Further, there are now four $\mu^{\pm}e^{\mp}2j$ events
available for analyses, and it will be valuable to compare the top mass
estimates from them with those from the $l^{\pm}4j$ events. That the two
groups of top mass estimates should be in agreement is a powerful test of
the methods now being used for top mass determinations.

\vspace{.25in}
\section{The Data and Analyses of ``$\mu^{\pm} e^{\mp} 2 jets$'' events}

  Over several years, we \cite{dalitz1,dalitz2,gold} and
K.~Kondo, {\it et.al.\/} \cite{kondo1,kondo2},
independently, have
developed a method for determining whether events of the type reported
by the CDF and D0 Collaborations are consistent with the hypothesis
\begin{equation} \bar{p} + p \rightarrow \bar{t} + t + {\rm other \: hadrons},
\label{pbarp}
\end{equation}
followed by
\begin{equation} t \rightarrow W^{+} b, \label{twb}
\end{equation}
where
\begin{equation} W^+ \rightarrow l^+\nu_l \:{\rm or}\: u\bar{d} \:{\rm or}
\: c\bar{s}, \label{wplus}
\end{equation}
and
\begin{equation} \bar{t} \rightarrow W^{-} \bar{b}. \label{tbarwb}
\end{equation}
where
\begin{equation} W^- \rightarrow l^-\bar{\nu}_l \:{\rm or}\: d\bar{u}\: {\rm
or}
\: s\bar{c}. \label{wminus}
\end{equation}
The procedure is to take the measured configuration of momenta for the
final leptons and jets in a single event and to evaluate the probability
$P(m_t)$ that these production and decay processes could produce the
observed configuration for an assumed value of the top quark mass.
This evaluation takes into account each step in the processes (1) and
(2): the parton content of the incoming proton and antiproton; the
differential cross section for these partons $q$ and $\bar{q}$ to produce
the $\bar{t} t$ pair; the decay probability for the top and antitop
quarks to undergo the processes~\ref{twb} and~\ref{tbarwb} respectively,
with the angles observed in the $t$ and $\bar{t}$ rest frames; and the
probabilities for the W bosons to decay into the leptons or quark pairs as
observed, retaining the spin and tensor polarization the
W bosons have following $t$ or $\bar{t}$ decay.

   We applied this procedure to the published dilepton event (which we
refer to as CDF-1) from the 1988-89 run by CDF\cite{sliwa,cdf3}. The
primary muon
from this event did not go through the muon detector which was in place for
that
first run, but its muonic character was made quite clear by the evidence
from its passage through an adjacent calorimeter. Since one of the
acceptance criteria in that run was that the muon should identify itself
by passing through the muon detector, CDF has rejected this event. To
avoid such rejection in subsequent runs, CDF has elongated their muon
detector, so that an event with the same configuration as CDF-1 would
be accepted today. This rejection is undoubtedly sound
practice when one is concerned with cross sections and relative
rates, but our only concern here is whether or not this event is a
satisfactory candidate for interpretation as an example of top-
antitop production and decay.  Our analysis of this event led to a
somewhat low mass value, either $121^{+18}_{-8} ~GeV$ or
$131^{+22}_{-12} ~GeV$, depending on
the details of the treatment of the transverse momentum
distribution of the parton-antiparton pair which generates the final
$t \bar{t}$  state.  A similar calculation has been made
independently by Kondo et al.\cite{kondo1,kondo2}, leading to a mass
distribution
$P(m_t)$ almost identical with our distribution with a peak at
$121 ~GeV$.  The D0 collaboration applied three modified
versions of our method to their $\mu e 2jet$ candidate and
obtained a probability distribution which peaks at $141 ~GeV$, with
half-maximum values at about $131 ~GeV$ and $165 ~GeV$.  The input
data for this event are not known to us.  The D0 detector does
not have the capability to measure the charge sign for $e^{\pm}$
but the charge sign for the muon can, in principle, be determined
by the magnetic field in the muon detector.

     The momentum data for the two $\mu^{\pm} e^{\mp} 2j$ events observed
by CDF\cite{cdf1,cdf2} in their 1992-93 run, mentioned above, are given here
in Table 1 in terms of Cartesian momenta.  We have analysed these
two events by the same procedure as we used\cite{dalitz1,dalitz2} for the CDF-1
event.  In both cases jet 3 is neglected for the reason given in
footnote 1.  We assigned uncertainties to the jet transverse
energies by an algorithm based on the $\sigma_E$ values
presented by CDF for all their different energy jets, in the single
lepton $t \bar{t}$ events\footnote{A linear interpolation, $\sigma(E) =
(3.4 + 0.1 E) GeV$ was used. This underestimates the uncertainties for some of
the low energy jets.}.
A jet is paired with a lepton to determine a paraboloid of possible
top momenta.  The second jet is paired with the other lepton to
form the paraboloid for the anti-top momenta.  Assuming that the
$t \bar{t}$ production occurs with small total transverse momentum, the
transverse momentum of the top quark (for a definite mass value
$m_t$) should nearly cancel the corresponding transverse momentum
of the anti-top quark with the same mass $m_t$.  To allow for
gluon bremsstrahlung, we took a gaussian distribution of
values centered on zero, with a representative width of $\rho =
0.1 m_t$, to weight the probability of the pair of transverse momenta
assigned to the $t$ and $\bar{t}$\cite{dalitz1}.
The resulting probability distributions are shown on Fig. 1.
In both cases, the assignment of jets to the $b$ and $\bar{b}$
quarks is uniquely determined by this mass analysis. It happens that for event
41722/38382, jets 1 and 2 are necessarily $b$ and $\bar{b}$,
respectively, and the same holds for event 41540/127085.  The
latter also has a secondary (slow) $\mu^{+}$ lepton, which would
routinely be assigned to the $\bar{b}$ jet, here jet 2. We discuss, in the
Appendix, the possible interpretation for this additional lepton.

     The probability distributions $P(m_t)$ from these two new
CDF events, shown in Fig. 1, peak at $158^{+7}_{-6} ~GeV$
for 41540/127085 and at $168^{+24}_{-14} ~GeV$ for
47122/38382. These independent probability distributions are
consistent with the production of top quarks with $158^{+7}_{-6} ~GeV$,
as seen in Fig. 2, where the two have been multiplied.  The
mass distributions for the CDF-1 and D0 events are
appreciably lower in peak mass, but each has a very gradual fall-
off for higher mass values.  The joint probability for all four
dilepton events is displayed in Fig. 2.  It peaks at $156^{+7}_{-6}
{}~GeV$,
since event 41540/127085 gives the narrowest distribution.
This event also has the lowest likelihood; indeed, if $\rho$, the
transverse momentum weighting parameter in eq.(3.21) of ref.\cite{dalitz1},
is reduced from our adopted value of $0.1 m_t$ down to zero, the
solution for this event ceases to exist.  However, as things are,
this joint probability $P(m_t)$ has a convincing shape,
although its absolute likelihood is very low\footnote{CDF-1
comes from an earlier run and the data for the D0 event is
not published, so we have not been able to normalize the probablities for
all the dilepton events consistently. The two recent CDF events are
consistently normalized however, so their likelihoods can be compared.}.
It is still quite
possible that not all of these events, or even any of them, are due
to top-antitop production and decay.  More candidate $e^{\pm} \mu^{\mp}
{}~2j$ events are needed for study before we can conclude which of these
events constitute a group of top-antitop events for a definite mass
$m_t$. As already discussed \cite{berends,dalitz1}, there are
non-top sources of
background possible which may mimic real top-antitop events.

\vspace{.25in}
\section{The Data on ``$l^{\pm} 4 jets$'' events and Analysis Procedure}

     We now consider the single lepton events, with at least 4
jets, one of which is tagged as $b$ or $\bar{b}$ quark.  The momenta of
the leptons and jets in the seven published events are given in
Cartesian co-ordinates in Table 2. The jet calorimeter energies
are the ``corrected values'' quoted by CDF\cite{cdf2} in their Appendix A,
following the calculated scatter plots given in their Fig. 57;
they are the CDF estimates for the original parton energies, with
well-defined statistical uncertainties.  The C.M. energy for
the two jets hypothesized to result from W decay is generally
rather far from the well-known value\cite{pdg} $M_W = 80.2(2) ~GeV$,
and this presents a problem. CDF uses a kinematic fitting procedure,
established long ago\cite{squaw} in bubble chamber work, to manipulate
the already
corrected transverse energies in order to reproduce the W mass value
at the expense of a higher $\chi^2$.
We do not use the resulting ``best fit values'' given by
CDF, since we follow a different scheme of analysis\cite{dalitz1}.

     Our analysis of these $l^{\pm}4j$ events employs a simple
extension of the method used above for dilepton events, which has
been laid out in considerable detail in ref. \cite{dalitz1}.  We sketch it
briefly here, for the case of a positively charged lepton $l^{+}$;
the case for $l^{-}$ follows when every particle is
replaced by its antiparticle and vice versa.  One jet is chosen
tentatively to be the $b_l$ jet associated\footnote{We use the CDF
notation\cite{cdf2}} with $l^{+}$ and a kinematic
paraboloid is formed, as before, leading to an ellipse in momentum space
which includes all momenta $\it{t}$ consistent with $\it{b}_l$ and
$\it{l}^{+}$, for an assumed mass $m_t$.  The other three jets
are assumed to arise from $\bar{t}$ decay where the resulting $W^{-}$
boson decays hadronically, thus:
\begin{equation} \bar{t} \rightarrow \bar{b} + W^{-},\:
 {\rm followed \: by}\:  W^{-}
\rightarrow \bar{q}_1 + q_2 . \label{wqq}
\end{equation}
The quark assumed to be $\bar{b}$ will be denoted$^5$ by $b_j$.
The experimental error distributions for these quark energies
have been discussed in much detail by CDF\cite{cdf2} and we adopt the same
algorithm that interpolates their $\sigma_E$ values as stated in footnote
3. A grid of momentum values $(\bar{\it{b}},\bar{\it{q}}_1,{\it q}_2)$
is laid out and weighted by their probability values at
each point, together with a probability weighting
$F_W(\bar{\it{q}}_1,\it{q}_2)$ of Breit-Wigner
form to emphasize those grid points at which ($\bar{\it{q}}_1 \cdot {\it
q}_2)$ is consistent with the W-boson mass.  At each grid point, there is a
definite momentum ($\bar{\it{t}} = \bar{\it{b}} + \bar{\it{q}}_1 +
\it{q}_2)$ and deduced mass $m_t$, and this point is then paired with
the points on
the t- ellipse for $m_t$, which also have their weighting factors due to
measurement errors.  This product of probabilities is
finally weighted by a Gaussian factor $G[(\it{t} + \bar{\it{t}})_T / \rho ]$
to represent the effect of limited transverse momentum due to initial
state gluon emission, the value $0.1 m_t$ being adopted for
the parameter $\rho$. Contributions to the net probability for the top quark
mass to lie within $(m_t , m_t + \Delta m_t)$ come from all grid points
which lie within this band $\Delta m_t$, and are summed to give the net
probability $P(m_t)$ indicated by this event.

     This probability has been computed for a definite assignment
$(b_l;\bar{q}_1,q_2,\bar{b}_j)$ of the four jets. If none of the jets is
b-tagged, then for an individual event distribution, we will still have to sum
over all permutations of the jet assignments.  If one jet is b-tagged
by SVX, we shall generally not know whether is is a $b$-jet or
a $\bar{b}$-jet, and we shall then have to sum over the two
possible assignments for a jet which may be due to $b$ or to $\bar{b}$.
It is important to note that each jet is treated in an
identical way, so that relative probability between different
events and different interpretations can be compared.  The probability
distributions obtained in
this way for each of the seven $l^{\pm}4j$ events are shown in Fig. 3.
In most, but not all, of the events, the combination of jets which
has the largest integrated probability is the assignment
chosen by CDF's method\cite{cdf2}.  However, two exceptions to this
observation will be mentioned below.  The product of the independent
probabilities for these seven $l^{\pm}4j$ events is plotted
in Fig. 2.  Their net distribution peaks at  $m_t = 172^{+2}_{-4} ~GeV$,
in accord with the CDF group's conclusion for the same seven events.

     How should we assess the significance of the integrated
probability found for each of these events?  They range from a
maximum of roughly $8 \times 10^{-4}$ for event 45880/31838 down to
$5 \times 10^{-7}$ for event 43351/266423.

\vspace{.25in}
\section{Our Analysis Procedure in the Light of a Monte Carlo Model for
$l 4 jets$ Events}

To address this question of significance, we have carried out the same
analysis procedure for a random sample of computer-generated events, a
Monte- Carlo simulation based on tree-level QCD Feynman graphs for
top-antitop production, followed by their decay into the final states
$l^{\pm}4j$. In this way, we generated 100,000 events for mass $m_t = 170
{}~GeV$, of which 38,394 passed all of the appropriate experimental cuts. Each
event requires the specification of eleven variables, each arranged to
vary over a finite range, and this finite 11-dimensional space was divided
into a finite number of cells\cite{barger}. The procedure is iterative,
based on a program by Ohnemus\cite{ohnemus}. Each event had a weight, its
fractional contribution
to the total cross section. The first 20,000 events, randomly generated,
led to a net weight in each cell, and a new set of variables were chosen,
leading to a new set of cells, each with about the same net weight. This
procedure was iterated five times, for each new set of 20,000
randomly-generated events. The distribution of the final 20,000 events in
the 11-dimensional space is then expected to be much closer to the
physical reality corresponding to this simple tree-level model, than that
of the first 20,000 events. The improvement in each iteration can be
tested, for example by comparing the total cross section calculated after
each iteration with the directly calculated total cross section for this
simple model, as given by Berends {\it et.al.}\cite{berends}. The weights
are finally
used to choose randomly a set of unweighted events to form a representative
subsample\cite{barger}. The subsample we can analyse is necessary small,
of order
100, since our analysis procedure is quite complicated and needs very
considerable computer time. In the end, we chose a subsample of 105 events
randomly, by the procedure just mentioned above, out of the 1,292 events
which passed the experimental cuts, from the first 3,000 events of the
last iteration. Our purpose is to compare the observed features of the
candidate top-antitop events with the features predicted for these events
by this simple QCD model.

Since these are MC events, we know which quarks are which. In analyzing
each event, typical measurement errors are assumed for the lepton and the
jets. We then use our analysis method to deduce the probability
distribution $P(m_t)$ from the event.
The integrated probabilities,
$IP\equiv\int dm_t P(m_t)$,  and the $m_t$
values for the peak in the probability $P(m_t)$ are shown in the ``scatter
plot'' of Fig.4. This shows that the analysis does reproduce the input mass
rather well, but that there is a wide range in the IP values. We have
plotted the distribution obtained for these IP values in Fig.5, and note
that they have a rather wide range. However, this exercise has given us
some criterion for recognizing poor fits, when we come to consider the
analysis of real events, and this was its purpose.

\vspace{.25in}
\section{The Analysis of Seven $l^{\pm} 4 jets$ Events}

Our analyses of the seven empirical events are summarized in
Table 3.  Since these are tagged events, the tagged jet is likely to be
either a $b$ or $\bar{b}$ jet associated with the $W$ that decayed either
leptonically (the $b_l$ jet) or hadronically (the $b_j$ jet). For a given
event, each assignment of jets to the $t - \bar{t}$ hypothesis in
eqn(\ref{wqq}) is analyzed independently, i.e. each particular
choice of jets to correspond to the configuration $(b_l; q_1, q_2, b_j)$,
is analyzed
separately.  CDF has done such separate analyses as well. They assign the
measured jets and the lepton to a configuration and determine the
$\chi^2$ per degree of freedom (there are 18 measurements - momenta, energies
and angles, with 20 constraints) that that configuration satisfies the
$t-\bar{t}$ hypothesis, based only on the jets' kinematics, without weighting
by the probabilities for such kinematics.

All but two of the seven events give integrated probabilities
between $4.3\times
{}~10^{-4}$ and $1.6\times ~10^{-4}$, significantly lower (by a factor about
1/5) than the majority of our Monte Carlo sample.  Two of these
events, 45880/31838 and 45879/123158, have two maxima peaking above
$10^{-4}$; the former gives  $4.3\times 10^{-4}$ with $m_t=164 ~GeV$ for
configuration $(b_l; q_1, q_2, b_j) = (1;2,4,3)$ and $3.2 \times
10^{-4}$ with $m_t=134 ~GeV$ for $(2;1,4,3)$, and the latter giving
$3.8 \times 10^{-4}$ with $m_t=178 ~GeV$ for $(2;3,4,1)$ and $2.5 \times
10^{-4}$ with $m_t=180 ~GeV$ for $(2;1,4,3)$. CDF's fit to the latter is
for configuration $(1;3,4,2)$ with $\chi^2 = 2.2$ and $m_t=169(10)
{}~GeV$; we do have a fit in this configuration with peak at $168 ~GeV$,
but it has a low integrated probability, $6 \times 10^{-6}$.
The events 43090/47223 and 43351/266423 also give much lower
integrated probabilities, $3 \times 10^{-6}$ and $5\times 10^{-7}$,
respectively, and we are inclined to reject them as top candidates.  The
latter is a poor fit also in CDF's analysis, but CDF found the former to
be acceptable.  On the other hand, event 45610/139604 gives
integrated probability $1.6 \times 10^{-4}$ with $m_t=180 ~GeV$ in
our analysis; the CDF analysis also found $m_t=180(9) ~GeV$ but
a poor $\chi^2 = 5.0$.       .

\vspace{.25in}
\section{Summary of Conclusions}

     We may summarize our considerations as follows:

1)   Our analysis of the seven $l^{\pm}4j$ events now known
is in general accord with the CDF-analysis, especially with their
mass estimate of about $175 ~GeV$. Two of the events have very low
likelihoods in our analysis, while two of them have relatively
large $\chi^2$ in the CDF analysis, one event being rejected by both;
four events stand firm in both analyses.  The three events rejected
may be due to background such as that originating from the
processes $W^{\pm}+4 jets$ with $W^{\pm} \rightarrow l^{\pm}$,
as discussed by Berends, $\it{et.al.}$ \cite{berends}, although those
authors show that tagging a single $b(\bar{b})$ quark should significantly
reduce that background. Their calculations indicate a suppression by about
$10^{-2}$ when both $b$ and $\bar{b}$ are tagged.  More estimates from
other mechanisms involving b-quarks need to be considered
quantitatively, within the framework of our analysis procedure.

(ii)  The relative rate between $l^{\pm}4jet$ and $e^{\pm}\mu^{\mp}2jet$
events must be considered. Accepting that four b-tagged events
of the former class have been observed, we need to calculate the expected
number of events of the latter class.  This is a complicated calculation,
which is sensitive to the precise cuts which are imposed and which
we do not attempt to carry out here.  The efficiencies depend on
whether the lepton in question is an electron or a muon.  The
nature of the identification given by tagging is different for SVX
and SLT.  SVX does not distinguish $b$ from $\bar{b}$, since it determines
only the location of the secondary vertex, while SLT does not give
the location of the vertex but does distinguish between $b$ and
$\bar{b}$. Since the c-quark decay lifetime is
shorter than that for the b-quark, there should frequently be seen
a tertiary vertex arising from c decay, not far from a secondary b-
vertex.

iii) Finally, the peak mass $m_t$ appears to be systematically somewhat lower
for $e^{\pm}\mu^{\mp}2jet$ events than for $l^{\pm}4jet$
events, the former being $156(8) ~GeV$, and the latter being $175(8)
{}~GeV$, a separation of $19(11) ~GeV$, although this is still a tolerable
difference.  We repeat
that it is far from sure that all of the latter events are due to $t
\bar{t}$ production and decay. However, to strengthen the lower mass
value for the former, we can also consider the CDF event 45047/104393,
which has a more natural
identification with the dilepton variety. Assigning
both leptons as ``hard'', rather than taking one as ``soft'', and combining
two jets into a single jet, leads to a
good fit as a $t \bar{t}$ event with $m_t=136^{+18}_{-14}$, consistent
with the generally lower masses for the dilepton candidates.

\vspace{.25in}
\section{Acknowledgements}

The authors are grateful to J. Ohnemus for supplying a copy of
an efficient Monte Carlo code with importance sampling and to P. Sphicas,
Tony Smith, K. Kondo, and D0 group members for many useful conversations.
G.R.G. thanks the U.S. Department of Energy for partial support of this
research (DE-FG-02-92ER40702), and Prof. John Negele for hospitality at the
MIT Center for Theoretical
Physics during a sabbatical leave when this work was begun. We appreciate
the hospitality of Prof. D. Sherrington at the Department of Theoretical
Physics, Oxford.

\vspace{.25in}
\appendix{\bf APPENDIX}

The CDF dilepton event 41540/127085 has a unique interpretation when
analysed as $\mu^+e^-j(1)j(2)$,  $j(1)$ being identified as the $b$-jet and
$j(2)$ as the $\bar{b}$-jet (see Table 1). The jet $j(3)$ is close to the
initial direction, being most probably due to gluon bremsstrahlung. The
event has a third lepton, a ``slow'' $\mu^+$
of energy $8.9 ~GeV$. The routine choice of associating this lepton with
the $\bar{b}$ jet is
not convincing since its largest momentum component, $p_x(\mu^{+}) = 8.7
{}~GeV$, is oppositely directed to the largest component of the jet 2
momentum, $p_x = -50.0 ~GeV$.
It is much more plausible that the slow $\mu^+$ is associated with jet 1,
since its momentum is almost parallel with the momentum of jet 1, and in
the same direction; its momentum $p_\perp$, transverse to jet 1 is only
about $0.6 ~GeV/c$.

The sequence of quark processes which lead to the emission of a tertiary
$\mu^+$ lepton, (a) with, and (b) without, a secondary $\mu^-$ lepton, are
as follows:
\begin{equation} t \rightarrow b + W^{+} \hspace{2in} W^{+}\rightarrow
\mu^{+}+\nu_{\mu}, \label{3a}
\end{equation}
\begin{equation} \hspace{0.5in} b \rightarrow c +W^{-} \hspace{1.2in}
W^{-} \rightarrow \mu^{-}+\bar{\nu}_{\mu} \label{3b}
\end{equation}
\begin{equation}
\hspace{1.5in} {\mathrm{or}}
\hspace{0.5in}
W^{-} \rightarrow {\mathrm{hadrons  (\bar{u} d + \bar{c} s)}}, \label{3c}
\end{equation}
\begin{equation} \hspace{1.0in} c \rightarrow s + W^{+} \hspace{1.0in} W^{+}
\rightarrow \mu^{+} + \nu_{\mu}, \label{3d}
\end{equation}
the W's in (\ref{3b}), (\ref{3c}) and (\ref{3d}) being necessarily virtual,
of course.
The net process for the event 41540/127085 would be the consequence of
(\ref{3a}),(\ref{3c}) and (\ref{3d}), thus:
\begin{equation} t \rightarrow {\mathrm{hadrons\/}} +
\mu^{+}({\mathrm{fast}}) + \mu^{+}({\mathrm{slow}}) + {\mathrm{neutrinos}}.
\label{4}
\end{equation}

To orient ourselves concerning the final states, we have made some simple
model calculations for the momentum distributions for a secondary lepton
$l_2$, or for  a tertiary lepton $l_3$, appropriate for an initial $b-$jet
with momentum about $130 ~GeV/c$. We adopted the fragmentation function of
Peterson \textit{et.al.}\cite{peterson}, with the parameter
$\epsilon_Q=(0.49)/m_Q^2  ~GeV^{-2}$, where $m_Q$ denotes the appropriate
heavy quark mass. In the first step, the $b$-quark generally undergoes
hadronization to a final ground state meson $B_u^{-}$, $B_d^{0}$ or
$B_s^{0}$ with spin-parity $0^{-}$, together with some number of light
mesons; we neglect explicit mention of hadronization to $\Lambda_b$
baryons, since such final states contribute much less and do not affect
the over-all conclusions. Using the standard model expression for the
momentum distribution of the lepton resulting from $b\rightarrow c l_2
\nu_l$ in the B-meson rest frame, we obtain the $l_2^{-}$ momentum
distribution in the lab frame by integrating this distribution over the
B-momentum distribution given by the fragmentation function. The
resulting energy distribution for secondary leptons is given in Fig.6(a).
We note that these energies run up to very large values. The mean
$l_2^{-}$ energy is $\approx 33 ~GeV/c$ and $50\%$ of the leptons have
energy greater than $29 ~GeV$. The distribution for $p_{\perp}$, the secondary
muon momentum transverse to the B-momentum, is given in Fig.7(a),
although we must note that the B-momentum is affected by the gluons and
light mesons emitted so that it differs a little from the $b-$jet axis
observed. The most probable value for $p_{\perp}(l_2)$ is $1.4 ~GeV/c$; its
median value is $1.35 ~GeV/c$. For $90\%$ of the secondary leptons,
$p_{\perp}(l_2)$ exceeds $0.6 ~GeV/c$; $70\%$ of them have $p_{\perp}(l_2) \geq
1 ~GeV/c$.
For tertiary leptons, we must first carry out the same calculation for
the lab momentum distribution of the c-quarks from the b-jet. Naturally,
this distribution is quite different from that for the secondary leptons,
because of the large mass value for the c-quark. The lab energy
distribution for the $l_3^{+}$ lepton from $c\rightarrow s l_3^{+}\nu_l$
decay is then obtained by integrating the latter, as given by the
standard model, over the fragmentation function for the ground state
$0^{-}$ ($D_u, D_d \:\mathrm{and}\: D_s$)-mesons from the c-jet distribution
just calculated. The resulting energy distribution for the $l_3^{+}$
lepton is shown in Fig.6(b). We note that these energies are much less
than those for secondary leptons but their distribution is very
asymmetric; their peak value is $\approx 0.5 ~GeV$, while their median
value is $\approx 5 ~GeV$. Above $1 ~GeV$, the distribution falls gradually
with increasing energy $E(l_3)$, by a factor of 3 from 5 to 15 GeV, and
then faster beyond; $25\%$ of the tertiary leptons have $E(l_3)\geq 10
{}~GeV$, but only $9\%$ have energies exceeding $15 ~GeV$. The distribution for
the transverse momentum $p_{\perp}(l_3)$ is shown on Fig.7(b). It peaks at
$0.35 ~GeV/c$ and is a little asymmetric; about $30\%$ of the events have
$p_{\perp}(l_3) \geq 0.6 ~GeV/c$, about $5\%$ have $p_{\perp}(l_3) \geq 1.0
{}~GeV/c$.

We now return to the consideration of event CDF-41540/127085. That the
``slow'' $\mu^{+}$ lepton is associated with jet 1
is supported by a close examination of the event shown in Fig.10 and
Table VII of the CDF paper \cite{cdf2}. There is a displaced vertex shown
in the SVX detector, at
$\bar{r}=0.33 ~cm$ from the origin of the event.
Comparison of the $\phi$ distribution in their Fig.10(b) with the entry
in their
Table VII shows us that the secondary vertex shown is associated with the b-jet
(jet 1). We are not told where the $8.9 ~GeV/c$ $\mu^+$ emerged.
The two most immediate possibilities are:

(T$_1$) the displaced vertex is a non-leptonic b-decay, the ratio
$\bar{r}/\bar{d}_B$ being 0.26, where $\bar{d}_B = \gamma_B c \tau_B$
is the mean distance of travel by the b-quark before decay,
$\gamma_B m_B = 131 ~GeV/c$, and $\tau_B$ being the B-meson lifetime.
The resulting c-quark then undergoes decay $c \rightarrow s \mu^{+}
\nu_{\mu}$, leading to a ``slow'' $\mu^+$ which is tertiary. The chance
that this c-decay occurs outside the SVX region is about
$e^{-((d-\bar{d}_B)/\bar{d}_D)}$
where $d=0.5 ~cm$ and $\bar{d}_D =\gamma_D c \tau_D$. Taking $\gamma_D$ to
have value about the same as $\gamma_B$, we then have $\bar{d}_D$ about
$0.37 ~cm$, which gives the chance of the c-quark escaping without detection
to be about $40\%$.

(T$_2$) The vertex observed is a tertiary decay, the slow
$\mu^+$ being
one of the tracks observed (whether or not it is identified) and coming from
the transition $c \rightarrow s\mu^{+}\nu_{\mu}$. The only question is
``where is the b-quark decay vertex?'' To give rise
to what is observed, there should then be a b non-leptonic vertex between the
origin and the displaced vertex, but perhaps so close to the tertiary
vertex, in view of the rapidity of c-decay relative to b-decay, that it may
be difficult to separate the two vertices. Also, in this case, there should
necessarily be a $\mu^+$ emitted from the displaced vertex, although there
is no clear record of this
$\mu^+$ in the SVX data. It is
difficult to estimate the probability for this outcome, without more detailed
information. A much closer examination of the SVX data on this event is needed.

Such tertiary leptons will not be rare.
The branching fraction (BF) for all leptonic modes is known
\cite{pdg} to be about $21.0(4)\%$ for the b quark and about $23(3)\%$, on
average, for the c quark\footnote{For the $D$ mesons, the BF's are
$34.4(38)\%$ for $D^+$ and $17.7(24)\%$ for $D^0$. From their known
lifetimes their leptonic decay rates are therefore
$3.3(4)\times~10^{11} ~s^{-1}$ and $4.3(7)\times~10^{11} ~s^{-1}$,
respectively, in fair agreement with each other. The well known inequality
between their total decay rates (and therefore between their leptonic
BF's) is due to a suppression of the non-leptonic decay modes of $D^+$
relative to those for the $D^0$. However, it is leptonic BF's which are
relevant for discussing the possibilities for tertiary leptons. The leptonic
BF's are not known for $D_s^+$, only the upper limit, $<20\%$, but its
total lifetime is within three standard deviations of that for $D^0$.},
assuming that the configurations
$(\bar{u} c), (\bar{d} c) \,{\mathrm{and\/}}\, (\bar{s} c)$ are produced
equally often.
Neglecting corrections for the efficiencies for detecting SLT's, generally
stated to be about $30\%$ but which may be substantially lower than
this for the detection of tertiary leptons, we may estimate that
the frequency of tertiary leptons without any secondary lepton is
comparable with the frequency of secondary leptons without any
tertiary lepton.

However, there is an alternative interpretation possible for the ``slow''
$\mu^{+}$:

(S$_d$ or S$_s$) The hadronization of the b-jet may lead to a charged
$B^{-}$ meson or to a neutral meson, $B_d^0$ or $B_s^0$. In the latter
two cases, the meson may
undergo the process of $(B_d^0,\bar{B}_d^0)$ mixing or $(B_s^0,\bar{B}_s^0)$
mixing,
and can then emit a $\mu^+$ lepton from the secondary process
$\bar{b} \rightarrow \bar{c} \mu^+ \nu_{\mu}$, then possible from the
$\bar{b}$-quark in the $\bar{B}_d^{0}$ or $\bar{B}_s^{0}$ components of
the final mixed ($\bar{B}^{0},B^{0}$)-meson state.
{}From data on b-jet development following the much studied process
$Z^0 \rightarrow b\bar{b}$, it is
known that the secondary $\mu^+$'s from this source have an intensity
$13\%$ of the total from the secondary ($\mu^{+}+\mu^{-}$) leptons from
the  initial
b-quark. These secondary $\mu^+$'s from mixing will have the same energy
spectrum as the $\mu^{-}$ secondary leptons from all three kinds of final
B-meson, which we have estimated from our model calculation to have the
form shown in Fig.6(a), a spectrum much harder than our estimate for the
tertiary $\mu^{+}$ spectrum, given in Fig.6(b).

We may now use these calculated probabilities curves to assess the
relative likelihood of the two hypotheses, T and S, just discussed above.

(S)  $b\rightarrow \bar{b}\rightarrow \bar{c} l^{+}$ and $b\rightarrow c
l^{-}$.

As noted above, it is known \cite{pdg} that the rate for $l^{+}$ is
$\epsilon = 0.13$ times that for ($l^{+}+l^{-}$) when the sum is over
$B_d^{0}$ and $B_s^{0}$ mesons. We denote the distribution of the final
secondary lepton by $P_2(E_l)$, shown in Fig.6(a), and the distribution of
the secondary lepton momentum transverse to the b-jet axis by
$Q_2(p_{l\perp})$, shown in Fig.7(a). From ref.\cite{pdg}, we take
$B_{bl}=0.207$ for the branching fraction $(b\rightarrow \mathrm{all} \:
l^{\pm})/(\mathrm{all\: b\: decays})$. The net rate for $l^{+}$, occuring as
secondary leptons, is given by
\begin{equation}
R_S = \epsilon\cdot B_{bl}\cdot P_2(E_l)\cdot Q_2(p_{l\perp}).
\label{5}
\end{equation}
per inital b quark.

(T)  $b\rightarrow c \rightarrow l^{+}$, with no secondary lepton.

Here we ignore the SVX detector, i.e. we do not require the second decay
to be visible within it. From ref.\cite{pdg}, we take $B_{cl}=0.34$ as the
branching fraction ($c\rightarrow \mathrm{all}\: l^{\pm})/(\mathrm{all\: c\:
decays})$. The net rate for $l^{+}$ is now,
\begin{equation}
R_T = (1-B_{bl})\cdot B_{cl}\cdot P_3(E_l)\cdot Q_3(p_{l\perp}).
\label{6}
\end{equation}
per initial b quark.

For the event of interest, we have $E_{\mu}=8.9 ~GeV$ and
$p_{\mu\perp}=0.60 ~GeV/c$. The interpretation S that the ``slow'' $\mu^{+}$
lepton is due to ($\bar{B}^{0},B^{0}$) mixing gives the rate per initial b
quark as
\begin{equation}
R_S=0.22\times 0.13\times 0.018\times 0.36 = 4.4\times 10^{-4}.
\label{7}
\end{equation}
With the tertiary interpretation T, we have the rate
\begin{equation}
R_T=0.78\times 0.33\times 0.035\times 1.24 = 1.12\times 10^{-2}.
\label{8}
\end{equation}

Hence the calculations with our simple model for the decay sequences
$b\rightarrow c l^{+}\nu$ and $b\rightarrow c\rightarrow s l^{+} \nu$
indicate that the likelihood that this $\mu^{+}$ is tertiary relative to
the likelihood that it is secondary - but results from ($\bar{B}^{0},B^{0}$)
mixing - is 25:1. The main factor depressing the rate $R_S$ is the low value
for $\epsilon$; surprisingly, the observed values for $E_l$ and
$p_{l\perp}$ do not distinguish clearly between the possibilities S and T.

It is of interest to compare event 41540/127085 with those SLT
$l4\mathrm{jets}$ events reported by CDF, which can be assigned uniquely
and kinematically to secondary lepton emission. These are the events where
the lepton charge has sign in accord with the decay $b\rightarrow c l^{-}
\bar{\nu}$ (or $\bar{b}\rightarrow \bar{c} l^{+} \nu$, for events which
stem from $\bar{t}$ production and decay). We note that the primary lepton
energies $E_{1l}$ have a reasonable spread of energies, from $24.7$ to $117.3
{}~GeV$, as shown in Tables 1 and 2. The two energies $E_{1l}$ in the dilepton
events lie at energies in the range $\approx 30-70 ~GeV$. The four ``slow''
leptons available have energies $E_{2l}$ whcih range from $2.4$ to $14.3 ~GeV$,
while the $\mu^{+}$ energy in event 41540/127085 lies in the middle of
this range. The same holds for its $p_{l\perp}$ value. Since the b-jet
energy in this event has a surprisinglly large value, ($E_{bj}+E(SLT)$)
being $\approx 141 ~GeV$ (overlooking the unknown neutrino energy
resulting from this b-decay), we might look instead at the weighted
energies $E(SLT)/(E(SLT)+E_{bj})$ and transverse momenta
$p_{l\perp}/(E(SLT)+E_{bj})$, listed in Table 4. Even then, their values
for 41540/127085 still lie within the ranges obtained for these parameters
from the four $l4\mathrm{jets}$ events. None of these numbers mark out this
SLT event as being obviously different from the other SLT events, except
for the charge sign for the ``slow'' $\mu^{+}$ and the magnitude of the
ratio $R_T/R_S$ discussed above.

Finally, we must compare these four SLT $l4\mathrm{jets}$ events with the
calculated spectra for our simple
($b\rightarrow c l^{-}\bar{\nu},\: \bar{b}\rightarrow \bar{c} l^{+} \nu$)
model. Figure 6a shows that the median value
predicted for secondary lepton energy $E_{2l}$ lies at $\approx 30 ~GeV$ for
the dilepton event 41540/127085, with $130 ~GeV$ for jet 1 lab energy
$E_{bj}$. The four SLT events have lower b-jet energies, ranging
from $39$ to $97 ~GeV$. This does not
effect the $p_{l\perp}$ spectrum, but it alters the $E_{2l}$ spectra. For
each event the energy spectrum
will depend on the boost from the decaying B rest frame to the lab frame
(in which the B meson is a fragment of the b-jet).
For the event with the lowest associated b-jet energy, event 43351, the
corresponding spectrum will have a median of about $9 ~GeV$, compared to
the measured $E_{2l} = 2.37 ~GeV$. The median energy grows roughly linearly
with jet energy, so
the secondary lepton in each of these four events have generally
lower energy, $E_{2l}$, than the predicted median.
Three of these SLT events have
$p_{l\perp}$ values that lie below $0.5 ~GeV/c$, whereas our model predicts its
median to be $\approx 1.35 ~GeV/c$. The accord of these data with our
calculations is neither striking nor unfavourable. It may be that the cuts made
on the data by the experimenters, aimed at picking out any background
events, have a much larger effect on the predicted curves in Figs.6 and 7
than we have anticipated.

Double-tagging, the combination of the secondary vertex detector (SVX)
and the obsevation of secondary leptons (SLT) together should provide
a powerful means
for interpretation of the nature of individual events, without a full
dynamical analysis (which would at best be possible only rarely).
The above analysis of event 41540/127085 illustrates this point quite
strongly.

\newpage
\begin{center}
Table 1. Dilepton event data reported by CDF\cite{cdf2}. Comments on the
right are from CDF.
\end{center}
\vskip 0.1in
\begin{tabular}{|c|r|r|r|r|l|} \hline
       &   $p_x$ &  $p_y$&   $p_z$&$E(GeV)$&   \\ \hline\hline
\multicolumn{6}{l}{Run 41540  Event 127085}    \\ \hline
$e^-$  &  18.657 & 11.658&  20.731&  30.229&   \\
$\mu^+$&  46.089 & 11.491&   8.114&  48.188&   \\
$\mu^+$&   8.714 & -1.225&   1.593&   8.943& soft $\mu^+$  \\
jet 1  & 129.725 &-18.232&  14.439& 131.793& with $\mu^+$  \\
jet 2  & -49.968 &-34.988& -34.564&  70.112& with $e^-$    \\
jet 3  &  -9.740 & 24.107&-245.219& 246.593& backward anti-proton jet \\
                                               \hline
\multicolumn{6}{l}{Run 47122  Event 38382}     \\ \hline
$e^+$  &  45.859 & 21.384&  54.141&  74.105&   \\
$\mu^-$&  37.209 &  2.602& -30.191&  47.987&   \\
jet 1  & -66.981 &-52.331&  58.191& 103.010&  with $e^+$ \\
jet 2  &  24.993 & -7.167&  46.244&  53.052&  with $\mu^-$ \\
jet 3  &  17.303 & -4.962&-246.138& 246.795&  backward anti-proton jet \\
                                                \hline
\end{tabular}

\newpage
\begin{center}
Table 2. Single lepton data \cite{cdf2} with one jet tagged as a $b$-jet via
the Silicon Vertex Detector (SVX) or the emission of a Soft Lepton (SLT).
The $j1,...$ labels correspond to CDF's jet numbers.
\end{center}
\vskip 0.1in
\begin{tabular}{|c|r|r|r|r|l|} \hline
       &   $p_x$ &  $p_y$&   $p_z$&$E(GeV)$&   \\ \hline\hline
\multicolumn{6}{l}{Run 40758  Event 44414}     \\ \hline
$e^+$  & -94.313 &-50.113&  48.523& 117.306&    \\
jet  j1&  86.267 & 26.685& -21.881&  92.913&  SVX \\
     j2& -26.220 & 74.310&  23.996&  82.373&  \\
     j3&  46.052 & 47.417&  43.659&  79.217&  \\
     j4&  30.613 &-22.003&  76.790&  85.545&  \\ \hline
\multicolumn{6}{l}{Run 43096  Event 47223}     \\ \hline
$e^-$  &  21.753 & 21.093& -27.316&  40.795&   \\
     j1&  78.068 &100.425&   2.544& 127.225&  SVX  \\
     j2& -70.137 & 29.785& 137.091& 156.845&  \\
     j3&  10.642 &-66.960&  81.787& 106.235&  \\
     j4& -34.707 &-14.202& 138.856& 143.831&  \\ \hline
\multicolumn{6}{l}{Run 43351  Event 266429}     \\ \hline
$\mu^-$&  24.577 & -1.062&  -1.723&  24.660&  \\
     j1& 109.365 &-75.333& 195.687& 236.494&  \\
     j2& -85.879 &  6.159& -15.582&  87.499&  \\
     j3&  24.815 &-21.905&   7.680&  33.979&  \\
     j4&  -3.595 & 36.122&  14.128&  38.953&  SLT \\
$\mu^-$&  -0.605 &  2.032&   1.057&   2.369&  $p_{l\perp}=0.46$ \\ \hline
\multicolumn{6}{l}{Run 45610  Event 139604}     \\ \hline
$\mu^+$&  52.325 & 11.153&  -9.682&  54.369&  \\
     j1&  11.612 & 76.423& -58.639&  97.025&  SVX  \\
     j2& -13.843 &-71.064& -74.320& 103.755&  \\
     j3&   3.167 &-36.061& -77.935&  85.932&  \\
     j4& -19.286 & -9.042&   1.492&  21.352&  \\ \hline
\end{tabular}

\newpage
\begin{center}
Table 2. (cont'd)
\end{center}
\vskip 0.1in
\begin{tabular}{|c|r|r|r|r|l|} \hline
\multicolumn{6}{l}{Run 45705  Event 54765}     \\ \hline
$e^-$  &  12.221 & 54.445&  42.329&  70.038&  \\
     j1& -74.864 &-49.953&  81.137& 121.174&  \\
     j2& -51.229 &  4.189& -11.399&  52.649&  \\
     j3&  15.072 &-54.971&  41.817&  70.694&  SLT \\
     j4&  31.974 & -9.305& 113.471& 118.256&  \\
$e^+$  &   1.523 &-10.995&   8.984&  14.280&  $p_{l\perp}=1.62$ \\ \hline
\multicolumn{6}{l}{Run 45879  Event 123158}     \\ \hline
$\mu^+$&  52.586 &  4.746& -11.170&  53.969&  \\
     j1& -75.575 & 27.724&-197.545& 213.317&  \\
     j2&  45.871 &-84.446& -10.592&  96.682&  SVX \& SLT \\
     j3&  33.259 & 25.812&   5.488&  42.456&  \\
     j4& -36.642 & -4.359& -16.765&  40.530&  \\
$\mu^-$&   6.680 &-11.732&  -1.488&  13.582& $p_{l\perp}=0.27$ \\ \hline
\multicolumn{6}{l}{Run 45880  Event 31838}     \\ \hline
$e^-$  &  -5.942 &-25.106&   4.146&  26.131&  \\
     j1&  98.037 & -7.188& -19.791& 100.273&  \\
     j2& -26.071 &-55.037&  80.329& 100.804&  \\
     j3&  18.082 & 39.344&  16.853&  46.464&  SLT  \\
     j4& -22.051 &-13.776& -16.246&  30.658&  \\
$e^-$  &   0.838 &  2.440&   1.088&   2.800& $p_{l\perp}=0.27$ \\ \hline
\end{tabular}

\newpage

\begin{center}
{}~Table 3. Output from our analysis of 7 single lepton events. CDF's fitted
values\cite{cdf2} using a kinematical program\cite{squaw} are listed below
their preferred jet assignment.
\end{center}
\vskip 0.1in
\begin{tabular}{|c|r|r|c|r|r|l|} \hline
jets               & IP                &$m_t$            & Best Fit     &
             &      CDF $m_t$&CDF $\chi^2$       \\ \hline
($b_l,q_1,q_2,b_j$)&                   &                 &($x,\bar{x}$) &
$m(t\bar{t})$&               &                   \\ \hline
                   &\multicolumn{4}{l}{CDF fitted values appear below}
             &               &                   \\ \hline
\multicolumn{7}{l}{ Event 40758} \\ \hline
(j4,j2,j3,j1)      & $4.1\times10^{-4}$&$170^{+13}_{-9}$ &(0.395,0.201) &
507          &               &                   \\
                   &                   &                 &(0.481,0.177) &
526          & $172\pm{11}$  & < 0.1             \\
(j2,j3,j4,j1)      & $3.7\times10^{-6}$&$184$            &(0.363,0.211) &
498          &               &                   \\ \hline
\multicolumn{7}{l}{ Event 43096} \\ \hline
(j1,j3,j4,j2)      & $3.0\times10^{-6}$&$162^{+8}_{-4}$  &(0.521,0.139) &
484          &               &                   \\
                   &                   &                 &(0.529,0.152) &
511          & $166\pm{11}$  & 2.0               \\
(j4,j2,j3,j1)      & $5.3\times10^{-8}$&$224$            &(0.417,0.172) &
481          &               &                   \\ \hline
\multicolumn{7}{l}{ Event 43351} \\ \hline
(j2,j1,j3,j4)      & $5.0\times10^{-7}$&$160^{+12}_{-6}$ &(0.447,0.168) &
493          &               &                   \\
                   &                   &                 &(0.409,0.157) &
455          & $158\pm{18}$  & 6.1               \\ \hline
\multicolumn{7}{l}{ Event 40758} \\ \hline
(j2,j3,j4,j1)      & $1.6\times10^{-4}$&$180^{+7}_{-13}$ &(0.110,0.379) &
367          &               &                   \\
                   &                   &                 &(0.093,0.446) &
365          & $180\pm{9}$   & 5.0               \\
(j1,j3,j4,j2)      & $8.2\times10^{-6}$&$112$            &(0.107,0.396) &
369          &               &                   \\ \hline
\multicolumn{7}{l}{ Event 45705} \\ \hline
(j3,j1,j2,j4)      & $4.2\times10^{-4}$&$190^{+11}_{-14}$&(0.409,0.148) &
443          &               &                    \\
                   &                   &                 &(0.593,0.097) &
430          & $188\pm{19}$  & 0.4                \\
(j4,j1,j2,j3)      & $1.1\times10^{-5}$&$156$            &(0.388,0.135) &
411          &               &                    \\ \hline
\multicolumn{7}{l}{ Event 45879} \\ \hline
(j2,j3,j4,j1)      & $3.8\times10^{-4}$&$179^{+12}_{-10}$&(0.140,0.406) &
423          &               &                    \\
(j2,j1,j4,j3)      & $2.5\times10^{-4}$&$180$            &(0.131,0.452) &
438          &               &                    \\
(j1,j3,j4,j2)      & $5.4\times10^{-6}$&$168$            &(0.130,0.273) &
396          &               &                    \\
                   &                   &                 &(0.129,0.420) &
419          &$169\pm{10}$   & 2.2                \\ \hline
\multicolumn{7}{l}{ Event 45880} \\ \hline
(j1,j2,j4,j3)      & $4.3\times10^{-4}$&$164^{+12}_{-10}$&(0.225,0.176) &
358          &               &                    \\
(j2,j1,j4,j3)      & $3.2\times10^{-4}$&$134^{+6}_{-8}$  &(0.295,0.358) &
358          &               &                    \\
                   &                   &                 &(0.312,0.132) &
365          & $132\pm8$     & 1.7                \\ \hline
\end{tabular}

\newpage

\begin{center}
{}~Table 4. Events available having a secondary or tertiary lepton. Energies
in GeV, momenta in GeV/c. Bracket denotes possible tertiary lepton.
\end{center}
\vskip 0.1in
\begin{tabular}{|c|r|r|r|r|r|} \hline
Event                        & 43351 &45705      & 45879     &
   45880          &    41540                     \\ \hline
$E_{1l}$                     &  24.7 &   70.0    & 54.0      &
   26.1           &   48.2                         \\ \hline
$E_{bj}$                     &  39.0 &   70.7    & 96.7      &
   46.5           &   131.8                      \\ \hline
$E_{2l}$                     &  2.37 &  14.28    & 13.58     &
   2.80           &   (8.9)                      \\ \hline
$E_{2l}/(E_{2l}+E_{bj})$ & 5.7\% &     16.0\%    & 12.6\%    &
   5.7\%          &   (6.8\%)                    \\ \hline
$p_{l\perp}$             &  0.46 &     1.615     &  0.27     &
   0.27           &    0.60                      \\ \hline
$p_{l\perp}/(E_{2l}+E_{bj})$ & 1.1\% & 1.45\%    & 0.25\%    &
   0.55\%         &   (0.43\%)                   \\ \hline

\end{tabular}

\newpage

\begin{center}
{\bf FIGURE CAPTIONS}
\end{center}
\vskip 0.1in
\noindent 1. P(m$_t$) plots (not normalized) for the published dilepton events.

\noindent 2. P(m$_t$) plots for the published events of type
``$l^{\pm}+4\mathrm{jets}$'' interpreted as
examples of $t\bar{t}$ production and decay.

\noindent 3. Combined P(m$_t$) plots for (i) the dilepton events, (ii) the
``$l^{\pm}+4\mathrm{jets}$''
events, and (iii) all of these events, taken together.

\noindent 4. Scatter plot for ``$m_t$ vs. Log.(integrated probablity)''
for events generated by Monte Carlo calculations, using QCD tree graphs
for $m_t=170 ~GeV$, using only
the right final configurations in analysing 100 randomly chosen MC events.

\noindent 5. The projection of the scatter plot of Fig.4 onto the
Log.(Int.Prob.) axis.

\noindent 6(a). The energy distribution in the Lab. frame, for the secondary
leptons resulting from the decay $b\rightarrow c+l^{-}+\nu_l$, for a b-quark
jet of initial energy $130 ~GeV$.

\noindent 6(b). The energy distribution in the Lab. frame, for tertiary
leptons resulting from the decay $c\rightarrow s+l^{+}+\nu_l$, for a b-quark
jet of initial energy $130 ~GeV$.

\noindent 7(a). The distribution of the momentum transverse to the b-jet axis,
for secondary leptons resulting from the decay $b\rightarrow
c+l^{-}+\nu_l$, for a b-quark jet of initial energy $130 ~GeV$.

\noindent 7(b). The distribution of the momentum transverse to the b-jet axis,
for tertiary leptons resulting from the decay $c\rightarrow
s+l^{+}+\nu_l$, for a b-quark jet of initial energy $130 ~GeV$.

\end{document}